\documentclass[letterpaper, 10 pt, conference]{ieeeconf}  

\IEEEoverridecommandlockouts                              

\usepackage{graphics} 
\usepackage{epsfig} 
\usepackage{mathptmx} 
\usepackage{times} 
\usepackage{amsmath} 
\usepackage{amssymb}  
\usepackage{multirow}
\usepackage{stfloats}
\usepackage[margin=19.1mm]{geometry}
\usepackage{ctable} 
\usepackage{siunitx} 
\title{\LARGE \bf
Residual Channel Attention Network for Brain Glioma Segmentation
}

\author{Yiming Yao$^{*1}$, Peisheng Qian$^{*2}$, Ziyuan Zhao$^{2}$, and Zeng Zeng$^{\dagger 2}$
\thanks{$^*$Contributed equally. $^{\dagger}$Corresponding author, email: zengz@i2r.a-star.edu.sg.
$^{1}$School of Microelectronics, Shanghai University, Shanghai, China, 200444. $^{2}$Institute for Infocomm Research, A*STAR, 1 Fusionopolis Way, 21-01 Connexis, Singapore 138632.}
}
\begin{document}
\renewcommand{\thefigure}{\arabic{figure}}

\maketitle
\thispagestyle{empty}
\pagestyle{empty}

\begin{abstract}
A glioma is a malignant brain tumor that seriously affects cognitive functions and lowers patients' life quality. Segmentation of brain glioma is challenging because of inter-class ambiguities in tumor regions. Recently, deep learning approaches have achieved outstanding performance in the automatic segmentation of brain glioma. However, existing algorithms fail to exploit channel-wise feature interdependence to select semantic attributes for glioma segmentation. In this study, we implement a novel deep neural network that integrates residual channel attention modules to calibrate intermediate features for glioma segmentation. The proposed channel attention mechanism adaptively weights feature channel-wise to optimize the latent representation of gliomas. We evaluate our method on the established dataset BraTS2017. Experimental results indicate the superiority of our method.
\newline
\indent \textit{Clinical relevance}— While existing glioma segmentation approaches do not leverage channel-wise feature dependence for feature selection, our method can generate segmentation masks with higher accuracies and provide more insights on graphic patterns in brain MRI images for further clinical reference.
\end{abstract}

\section{INTRODUCTION}
\begin{figure}[ht!]
    \centering
    \includegraphics[width = 0.8\linewidth]{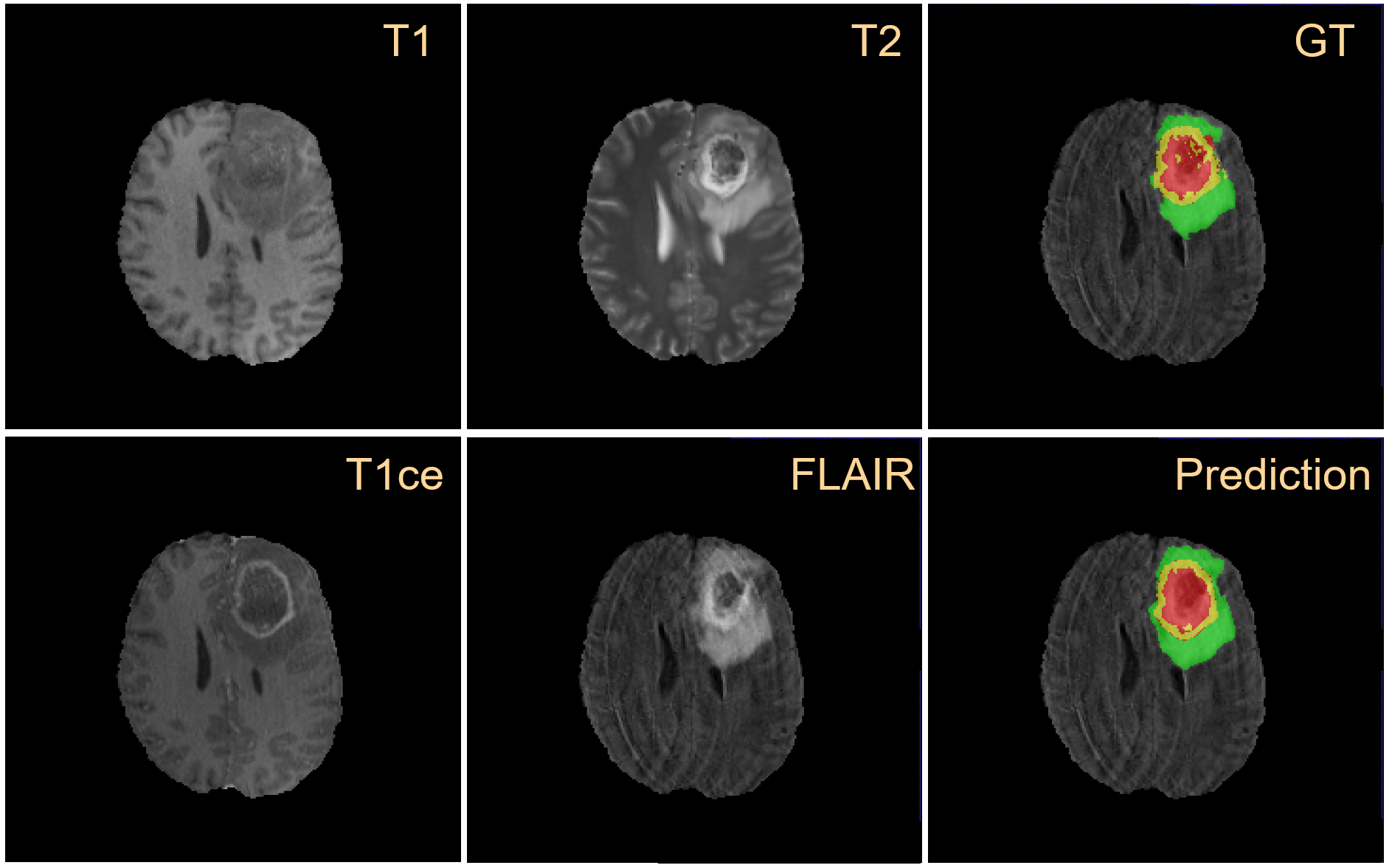}
    \caption{The BraTS2017 MRI image slices include 4 modalities (T1, T1ce, T2, and FLAIR). In ground-truth (GT) and predicted segmentation maps, the whole tumor (WT) includes all colors (red, yellow, and green). The tumor core (TC) includes red and yellow, and the enhancing tumor (ET) is red.}
    \label{fig:fig_1}
\end{figure}

Gliomas are among the most common and dreadful malignant brain tumors. It starts from non-neuronal glial cells and leads to symptoms including seizures, cranial nerve disorders, and ultimately, death~\cite{black2005cancer}. Diagnosis of brain glioma at early stages is essential for subsequent treatment and recovery, which can be performed in multi-modality brain Magnetic Resonance Imaging (MRI). The $4$ modalities in brain MRI includes T1, T1ce (T1 with contrast-enhanced), T2, and FLAIR (Fluid Attenuation Inversion Recovery). Segmentation of brain tumor, as demonstrated in Fig.~\ref{fig:fig_1}, is a crucial step of the diagnosis, which localizes the glioma for quantitative investigation~\cite{shin2012hybrid}. Manual segmentation requires expertise and labor, which is difficult to scale or satisfy the increasing demand for high-quality tumor segmentation.

Computer-aided prognosis has automated the process of brain tumor segmentation. Early works have approached the problem with classic machine learning approaches,~\emph{e.g.}, logistic regression~\cite{shin2012hybrid}, combinations of Bayesian classification and Markov Random Field~\cite{corso2008efficient}, and hierarchical approaches~\cite{bauer2012segmentation}. In recent years, deep learning methods such as deep convolutional neural networks (DCNNs) have been validated to be more effective in feature recognition on numerous biomedical applications, including disease diagnosis~\cite{zhao2019bira, qian2021two,deb2021trends}, health monitoring~\cite{yi2018enhance,bhatwearable, an2021mgait, an2021mars}, biomedical image analysis~\cite{ronneberger2015u, zhao2020deeply, 9629941, gu2020multi, zhao2021dsal}, and Electroencephalography (EEG) analysis~\cite{qu2020using,ding2021learning}. Nevertheless, the problem remains challenging in brain tumor segmentation for various reasons. First, voxel ambiguities between different tumor classes make it difficult to identify the clear boundaries of brain tumors~\cite{liu2021canet}. Moreover, generalization of tumor features is a demanding task, as there are noticeable variances in the shapes, sizes, and positions of gliomas in MRI~\cite{bauer2012segmentation}. Besides, it requires more effective feature extraction and selection to overcome inhomogeneity in MRI data.  

To effectively address the challenges mentioned above, we design a residual channel attention mechanism to calibrate channel-wise feature dependencies, by which the network learns to discriminate regions of brain gliomas. Our lightweight attention module includes 3D global maximum and average pooling on the hidden features' height, width, and depth. It embeds global channel-wise feature responses to be leveraged by the intermediate layers. We connect the attention modules at each level of an encoder-decoder segmentation architecture. Attention maps generated from attention modules are added to the hidden features to recalibrate the hidden features adaptively. The addition of hidden features also acts as residual connections in the attention module for model optimization. The Main contributions in our paper are summarized in the following,

\begin{figure*}[th]
\centering
\includegraphics[width=0.8\textwidth]{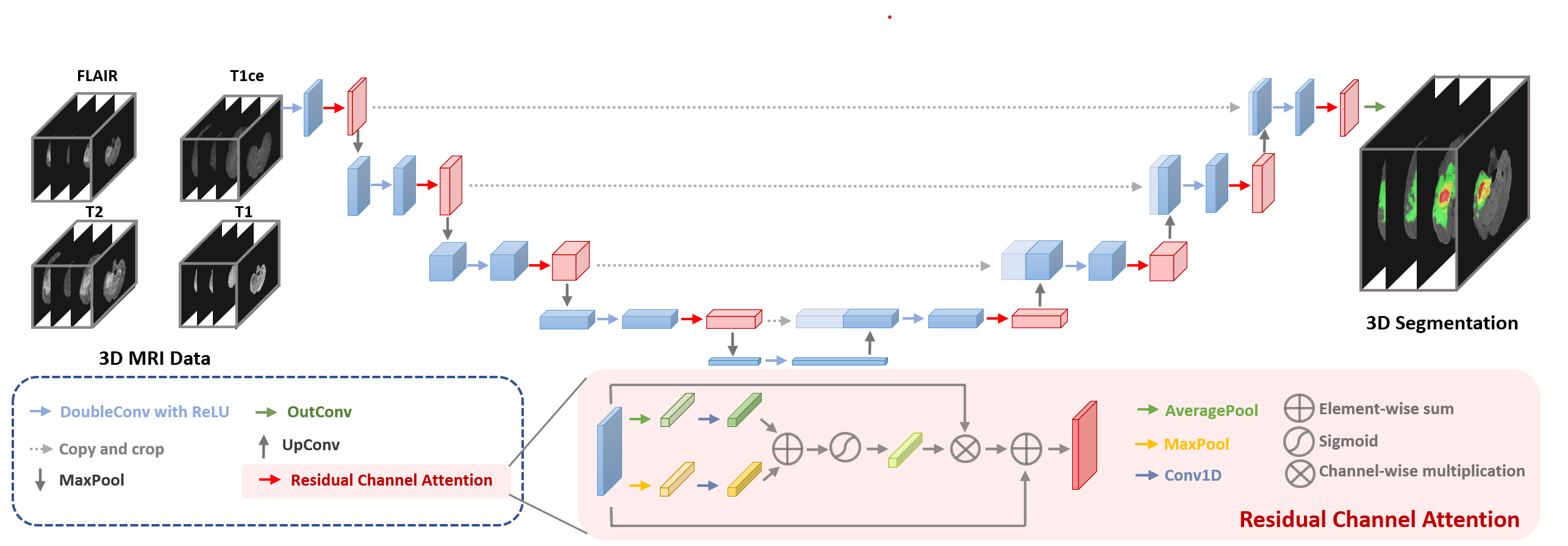}
\caption{The proposed segmentation architecture. Residual channel attention modules (labeled as red arrows and cubes) are appended after each DoubleConv layer. Derivation of channel descriptors and residual connections are illustrated in the enclosed rounded rectangle.}
\label{fig:arch}
\end{figure*}

\begin{itemize}
    \item We design a brain glioma segmentation architecture with a residual channel attention mechanism. Our proposed attention module dynamically calibrates intermediate features, which enhances the representational capacity of the network.
    \item We perform extensive experiments on the proposed attention module with various backbone networks and in conjunction with other feature selection and aggregation techniques. Experiments indicate the effectiveness of our approach with superior results than baselines. 
\end{itemize}

\section{Related Work}
\textbf{Brain glioma segmentation.} Traditional machine learning algorithms have been considered in brain glioma segmentation. A hybrid of clustering and logistic regression algorithm was developed to segment tumors and edema represented by sparse coding~\cite{shin2012hybrid}. Random decision forests were devised for voxel labeling in brain structures~\cite{festa2013automatic}. The Bayesian model was leveraged to improve affinity calculation, which was included in the segmentation by weighted aggregation (SWA) algorithm~\cite{corso2008efficient}. These methods are constrained by the representation capacity of hand-crafted features and require profound domain knowledge. DCNNs have achieved remarkable progress in brain glioma segmentation. Back in 2013, the possibility to apply DCNNs to brain glioma segmentation was investigated in~\cite{zikic2014segmentation}. Variations of 3D U-Net were proven effective in 3D graphical information extraction from brain MRI images~\cite{cciccek20163d, brugger2019partially, oktay2018attention}. Recently, the contextual information of tumors and surroundings were modeled in a feature interaction graph to optimize latent feature representation~\cite{liu2021canet}. Despite growing evaluation results, voxel ambiguity remains a challenging problem in brain glioma segmentation, which has yet to be solved. 

\textbf{Channel attention.} Channel attention mechanism adaptively recalibrates channel-wise features by exploiting dependencies between channels. A milestone in this field is the formulation of Squeeze-and-Excitation (SE) blocks~\cite{hu2018squeeze,zhao2020sea}. Efficient Channel Attention (ECA) was devised to alleviate dimensionality reduction in SE blocks while greatly reducing model complexity~\cite{wang2020eca}. Channel attention mechanism has been used in conjunction with DCNN for medical imaging segmentation~\cite{cciccek20163d, guo2021channel}, which inspires our work.

\section{METHODOLOGY}
\label{sec:METHODOLOGY}
The residual channel attention network is illustrated in Fig.~\ref{fig:arch}. 3D MRI data, including all modalities, are fed as input of the network. The network includes an encoder-decoder architecture modified from U-Net~\cite{ronneberger2015u}. The channel attention module, which is described in Sec.~\ref{sec:channel_attention_module}, is appended after each double convolution module during the encoding and decoding process. Intermediate features from convolution modules are calibrated by interdependence between channels modeled in attention modules, which is described in Sec.~\ref{sec:feature_calibration}. The output of the decoder consists of 3D segmentation maps for all tumor classes on the 3D MRI data.

\subsection{Channel Attention Mechanism}
\label{sec:channel_attention_module}
The proposed channel attention module is the core to discover channel-wise information and relationship in brain MRI data. For input feature $X\in R^{H\times W\times D\times C}$, $2$ feature descriptors,~\emph{i.e.}, the max-pooling and average-pooling results at channel $c$ can be calculated in the following,
\begin{equation}
    {X}_{max\_pool}=Max(X(h,w,d)), h\in [0, H), w\in [0, W), d\in [0, D),
\end{equation}
\begin{equation}
    {X}_{avg\_pool}=\frac{1}{H\times W\times D}\sum_{h=0}^{H-1}\sum_{w=0}^{W-1}\sum_{d=0}^{D-1}X(h,w,d),
\end{equation}
in which ${X}_{max\_pool}\in R^{1\times 1\times 1\times C}$ and ${X}_{avg\_pool}\in R^{1\times 1\times 1\times C}$. $H$, $W$, $D$ and $C$ represents height, width, depth, and channels of $X$, respectively. The $2$ descriptors are then transformed in a 1D convolutional layer $Conv$ with shared weights, which efficiently captures channel interaction without dimensionality reduction. Channel-wise addition and Sigmoid activation $\sigma$ are calculated on the output channel weights sequentially to acquire the combined channel attention map $M(X)\in R^{1\times 1\times 1\times C}$, as shown in the following Equ.~\ref{eqn:feature_map},
\begin{equation}
    M(X)=\sigma(Conv({X}_{max\_pool}) + Conv({X}_{avg\_pool})).
    \label{eqn:feature_map}
\end{equation}
The proposed module is inspired by Modified Efficient Channel Attention (MECA) in~\cite{guo2021channel}. However, there are $2$ substantial distinctions. First, there is one more dimension in the attention map in this paper, covering the depth and modalities of the data. Next, the features guided by the attention maps are passed to the subsequent Convolution layers, rather than to the skip connections in~\cite{guo2021channel}.

\subsection{Feature Calibration with Residual Connection}
\label{sec:feature_calibration}
We calibrate the intermediate feature $X$ with the attention map $M(X)$ in $2$ steps. First, we apply~\textbf{channel-wise} multiplication on $X$ and $M(X)$. Next, we perform~\textbf{element-wise} summation on the multiplication result and $X$. The calibrated feature $X^{\prime}$ can be calculated in Equ.~\ref{eqn:feature_calibration},
\begin{equation}
    {X^{\prime}} = (M(X)\bigotimes X)\bigoplus X,
    \label{eqn:feature_calibration}
\end{equation}
where $\bigotimes$ and $\bigoplus$ are channel-wise multiplication and element-wise summation, respectively. The summation includes a residual connection from the start of the attention module, enhancing gradient propagation in training.

\begin{table*}[!t]
\caption{Experimental results on BraTS2017.}
\label{tab:results} 
\centering
\begin{tabular}{c|cccc|ccc|ccc|ccc} 
\toprule
Method              & \multicolumn{4}{c|}{Dice Score}                 & \multicolumn{3}{c|}{Sensitivity} & \multicolumn{3}{c|}{Specificity}                 & \multicolumn{3}{c}{Hausdorff95}                   \\ 
\cline{2-14}
                    & Mean           & ET             & WT    & TC    & ET    & WT    & TC               & ET             & WT             & TC             & ET             & WT             & TC              \\ 
\hline
Attention U-Net~\cite{oktay2018attention}     & 0.771          & 0.672          & 0.863 & 0.778 & 0.847 & 0.900 & \textbf{0.862}           & 0.996          & 0.990          & 0.992          & 9.347          & 9.676          & 10.668          \\
CANet-concat~\cite{liu2021canet}          & 0.782          & 0.682          & 0.861 & 0.803 & \textbf{0.857} & \textbf{0.922} & 0.861            & 0.997          & 0.989          & 0.994          & 7.755          & 9.377          & 11.432          \\
3D U-Net~\cite{cciccek20163d}            & 0.794          & 0.706          & 0.865 & 0.810 & 0.803 & 0.906 & 0.829            & 0.998          & 0.990          & 0.995          & 6.624          & 8.193          & 8.958           \\
PRU-Net~\cite{nuechterlein20183d}             & 0.805          & 0.710          & \textbf{0.891} & \textbf{0.814} & 0.788 & 0.900 & 0.841            & 0.998          & 0.990          & 0.996          & 7.205          & 7.414          & 9.187           \\ 
\hline
RCAN (CANet-concat)      & 0.798          & 0.724          & 0.883 & 0.788 & 0.765 & 0.906 & 0.830            & 0.998          & \textbf{0.998} & \textbf{0.997} & 6.439          & \textbf{7.093} & 7.941           \\
\textbf{RCAN (U-Net)} & \textbf{0.807} & \textbf{0.743} & 0.882 & 0.795 & 0.771 & 0.913 & 0.850            & \textbf{0.999} & 0.997          & \textbf{0.997} & \textbf{5.854} & 7.514          & \textbf{7.561}  \\
\bottomrule
\end{tabular}
\end{table*}

\section{EXPERIMENTS}
\label{sec:experiments}
\subsection{Dataset and Implementation}
\label{sec:dataset_implementation}
We conducted experiments on the public Multimodal Brain Tumor Segmentation Challenge 2017 (BraTS2017), consisting of $285$ instances. Each instance contains $4$ MR sequences, including T1, T1ce, T2, and FLAIR. The sequence's height, width, and depth are $240$, $240$, and $155$, respectively. Ground-truth annotations contain healthy tissues, enhancing tumor (ET), whole tumor (WT), and tumor core (TC). We followed pre-processing steps in~\cite{liu2021canet} and performed five-fold cross-validation with a random division.

The implementation details are described as follows. Multiple augmentations were applied, including random rotation between $\pm20$ degrees, random elastic deformation ($\sigma=10$), random flipping, random scaling, and intensity shift of up to $10\%$. The data were randomly cropped into $128\times128\times128$ before being fed into the network with batch size $b_s=1$. The network was randomly initialized without pre-training and was trained for $200$ epochs using Adam optimizer with an initial learning rate of $0.0001$. The weight decay was set to \num{1e-5}. The learning rate was multiplied by $0.2$ after epoch $100$ and $150$. The experiments were carried out on an NVIDIA RTX A5000 GPU with PyTorch 1.9.1.

\subsection{Evaluation Metrics}
We follow existing works~\cite{wang2017automatic, liu2021canet} and use dice score, sensitivity, specificity, and Hausdorff95 distance to measure the segmentation performance. In particular, a lower Hausdorff95 distance between prediction and ground-truth indicates better model performance, while a higher dice score/sensitivity/specificity is more desired. Metrics for the $3$ tumor regions,~\emph{i.e.},  ET, WT and TC are computed separately.

\subsection{Baseline Methods}
\label{sec:baseline}
Several baselines are compared to our framework. In~\cite{cciccek20163d}, the 3D variant of U-Net was proposed. Spatial attention was introduced to U-Net in~\cite{oktay2018attention}. Reversible blocks were integrated into U-Net to reduce memory cost in~\cite{nuechterlein20183d}. In~\cite{liu2021canet}, a feature interaction graph was combined with U-Net to raise context awareness and reduce local ambiguity. Our proposed Residual Channel Attention Networks (RCAN) are described as below, 

\begin{itemize}
\item RCAN (U-Net): 3D U-Net with residual channel attention integrated.
\item RCAN (CANet-concat): Residual channel attention integrated into the backbone CANet, with the concatenated feature interaction graph~\cite{liu2021canet}.
\end{itemize}

\subsection{Results and Discussion}
\label{sec:results}
In Table.~\ref{tab:results}, experimental results of our method and baselines are presented. Comparing 3D U-Net and RCAN (U-Net), our method achieves a better result under most metrics for all tumor classes. It thus proves that our residual channel attention can calibrate the features learned by the model and enhance its performance. Moreover, the proposed network with channel attention outperforms Attention U-Net~\cite{oktay2018attention}, which attempts to optimize the U-Net with spatial attention. Our method also achieves the lowest Hausdorff95 distance between the predicted segmentation masks and the ground-truth in ET (enhancing tumor) and TC (tumor core), indicating more accurate edge predictions in the segmentation maps. Besides, RCAN (U-Net) reaches higher specificity but lower sensitivity than other U-Net variants in ET and TC, which is a possible direction for future optimization. With CANet-concat as the backbone, we further improve the output of the original CANet-concat. The result indicates the applicability of the proposed channel attention in conjunction with other feature selection mechanisms. 

\begin{figure}[t]
    \centering
    \includegraphics[width = 8.5cm]{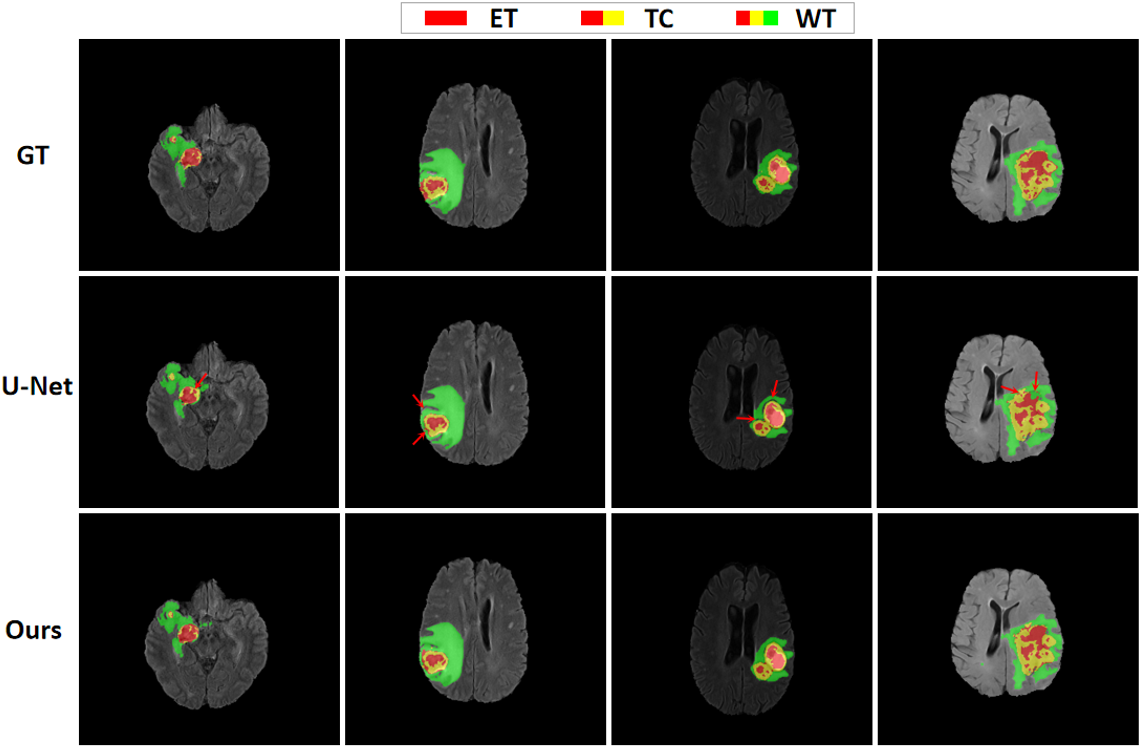}
\caption{Illustrations of segmentation maps from the ground-truth (GT), U-Net~\cite{cciccek20163d}, and our RCAN (U-Net) in BraTS2017. Arrows indicate incorrect segmentation in the comparison method.}
    \label{fig:visualization}
\vspace{5pt}
\end{figure}

In Fig.~\ref{fig:visualization}, we visualize selected segmentation maps. We observe that our network is capable of detecting small, isolated enhancing tumor (ET) areas, and can also predict the correct edges for ET, where the comparison method fails. Moreover, there are visible improvements in our edge prediction for tumor core (TC) and whole tumor (WT), which are in accordance with results from Table.~\ref{tab:results}.

\begin{table}[htbp]
\caption{Experimental results on the exclusion of components in residual channel attention.}
\label{tab:ablation}
\centering

\begin{tabular}{c | c c c c}
\specialrule{.1em}{.05em}{.05em} 
\multicolumn{1}{c}{} & \multicolumn{4}{c}{Dice Score} \\
\hline
Experiments & Mean & ET & WT & TC\\
\hline
Without max pooling & $0.796$ & $0.717$ & $0.889$ & $0.782$  \\
\hline
Without average pooling & $0.790$  & $0.715$ & $0.882$ & $0.785$ \\
\hline
Without residual connection &  $0.793$ & $0.688$ & \textbf{0.892} & $0.795$\\
\specialrule{.1em}{.05em}{.05em} 
\textbf{With all components} & \textbf{0.807} & \textbf{0.743} & 0.882 & \textbf{0.795} \\
\specialrule{.1em}{.05em}{.05em} 
\end{tabular}
\end{table}

We performed ablation studies on components in the residual channel attention network. The result is presented in Table.~\ref{tab:ablation}. The overall performance drops when any one of the pooling operations or the residual connection is removed. The performance decreases most when the average pooling operation is excluded, even slightly worse than the baseline U-Net. In summary, the ablation studies prove the necessity of each component in the proposed architecture.

\section{CONCLUSIONS}
In conclusion, we have developed a novel network for brain glioma segmentation. To address class ambiguities and high variability in tumor features, we devise a residual channel attention mechanism, which calibrates latent features and enhances the representational power of the network. Experiments are conducted on the 3D brain MRI dataset, BraTS2017. The results prove the effectiveness of the proposed network quantitatively and qualitatively. In the future, domain knowledge could be incorporated to the algorithm.

\addtolength{\textheight}{-12cm}   








\bibliographystyle{IEEEbib}
\bibliography{refs.bib}

\end{document}